\pdfoutput=1
\documentclass[12pt,a4paper]{article}

\usepackage{ifthen} 
\newboolean{pdflatex}
\setboolean{pdflatex}{true} 

\newboolean{articletitles}
\setboolean{articletitles}{true} 

\newboolean{uprightparticles}
\setboolean{uprightparticles}{false} 

\def\paperauthors{LHCb collaboration} 
\def\paperasciititle{Discovery potential of LHCb Upgrade II} 
\def\papertitle{Discovery potential\\of LHCb Upgrade II} 
\def\paperkeywords{ {LHCb}} 
\def\papercopyright{\the\year\ CERN for the benefit of the LHCb collaboration} 
\def\paperlicence{CC BY 4.0 licence}
\def\paperlicenceurl{https://creativecommons.org/licenses/by/4.0/}

\usepackage[top=1in, bottom=1.25in, left=1in, right=1in]{geometry}

\usepackage{microtype}
\usepackage{lineno}  
\usepackage{xspace} 
\usepackage{caption} 

\usepackage{graphicx}  
\usepackage{color}
\usepackage{colortbl}
\graphicspath{{./figs/}} 

\usepackage{amsmath} 
\usepackage{amssymb}
\usepackage{amsfonts}
\usepackage{upgreek} 

\usepackage{multirow}

\newcommand*\patchAmsMathEnvironmentForLineno[1]{%
\expandafter\let\csname old#1\expandafter\endcsname\csname #1\endcsname
\expandafter\let\csname oldend#1\expandafter\endcsname\csname
end#1\endcsname
 \renewenvironment{#1}%
   {\linenomath\csname old#1\endcsname}%
   {\csname oldend#1\endcsname\endlinenomath}%
}
\newcommand*\patchBothAmsMathEnvironmentsForLineno[1]{%
  \patchAmsMathEnvironmentForLineno{#1}%
  \patchAmsMathEnvironmentForLineno{#1*}%
}
\AtBeginDocument{%
\patchBothAmsMathEnvironmentsForLineno{equation}%
\patchBothAmsMathEnvironmentsForLineno{align}%
\patchBothAmsMathEnvironmentsForLineno{flalign}%
\patchBothAmsMathEnvironmentsForLineno{alignat}%
\patchBothAmsMathEnvironmentsForLineno{gather}%
\patchBothAmsMathEnvironmentsForLineno{multline}%
\patchBothAmsMathEnvironmentsForLineno{eqnarray}%
}

\usepackage[pdftex,
            pdfauthor={\paperauthors},
            pdftitle={\paperasciititle},
            pdfkeywords={\paperkeywords},
]{hyperref}

\usepackage{hyperxmp}

\hypersetup{pdfcopyright={Copyright (C) \papercopyright}}
\hypersetup{pdflicenseurl={\paperlicenceurl}}

\usepackage[bottom,flushmargin,hang,multiple]{footmisc}

\usepackage[all]{hypcap} 

\usepackage{xspace} 
\usepackage{upgreek}

\def\upgradetwo {\mbox{Upgrade~II}\xspace}

\def\MagUp {\mbox{\em Mag\kern -0.05em Up}\xspace}

\ifthenelse{\boolean{uprightparticles}}%
{
 
 \def\Pgamma      {\ensuremath{\upgamma}\xspace}

 \def\Pmu         {\ensuremath{\upmu}\xspace}                 
 \def\Pnu         {\ensuremath{\upnu}\xspace}                 
                  
 \def\Ppi         {\ensuremath{\uppi}\xspace}

 \def\Ptau        {\ensuremath{\uptau}\xspace}

 \def\Ppsi        {\ensuremath{\uppsi}\xspace}

 \def\PDelta      {\ensuremath{\Delta}\xspace}                 
 \def\PXi         {\ensuremath{\Xi}\xspace}                 
 \def\PLambda     {\ensuremath{\Lambda}\xspace}                 
 \def\PSigma      {\ensuremath{\Sigma}\xspace}                 
 \def\POmega      {\ensuremath{\Omega}\xspace}                 
 \def\PUpsilon    {\ensuremath{\Upsilon}\xspace}
 \let\oldPi\Pi
 \def\PPi         {\ensuremath{\oldPi}\xspace}

 \def\PB      {\ensuremath{\mathrm{B}}\xspace}                 
                  
 \def\PD      {\ensuremath{\mathrm{D}}\xspace}

 \def\PJ      {\ensuremath{\mathrm{J}}\xspace}                 
 \def\PK      {\ensuremath{\mathrm{K}}\xspace}

 \def\Pb      {\ensuremath{\mathrm{b}}\xspace}                 
 \def\Pc      {\ensuremath{\mathrm{c}}\xspace}                 
 \def\Pd      {\ensuremath{\mathrm{d}}\xspace}                 
 \def\Pe      {\ensuremath{\mathrm{e}}\xspace}

 \def\Pi      {\ensuremath{\mathrm{i}}\xspace}

 \def\Pp      {\ensuremath{\mathrm{p}}\xspace}

 \def\Ps      {\ensuremath{\mathrm{s}}\xspace}                 
 \def\Pt      {\ensuremath{\mathrm{t}}\xspace}                 
 \def\Pu      {\ensuremath{\mathrm{u}}\xspace}

 \def\thebaroffset{0.0em}
}
{
 
 \def\Pgamma      {\ensuremath{\gamma}\xspace}

 \def\Pmu         {\ensuremath{\mu}\xspace}                 
 \def\Pnu         {\ensuremath{\nu}\xspace}                 
                  
 \def\Ppi         {\ensuremath{\pi}\xspace}

 \def\Ptau        {\ensuremath{\tau}\xspace}

 \def\Ppsi        {\ensuremath{\psi}\xspace}                 
                  
 \mathchardef\PDelta="7101
 \mathchardef\PXi="7104
 \mathchardef\PLambda="7103
 \mathchardef\PSigma="7106
 \mathchardef\POmega="710A
 \mathchardef\PUpsilon="7107
 \mathchardef\PPi="7105
                  
 \def\PB      {\ensuremath{B}\xspace}                 
                  
 \def\PD      {\ensuremath{D}\xspace}

 \def\PJ      {\ensuremath{J}\xspace}                 
 \def\PK      {\ensuremath{K}\xspace}

 \def\Pb      {\ensuremath{b}\xspace}                 
 \def\Pc      {\ensuremath{c}\xspace}                 
 \def\Pd      {\ensuremath{d}\xspace}                 
 \def\Pe      {\ensuremath{e}\xspace}

 \def\Pi      {\ensuremath{i}\xspace}

 \def\Pp      {\ensuremath{p}\xspace}

 \def\Ps      {\ensuremath{s}\xspace}                 
 \def\Pt      {\ensuremath{t}\xspace}                 
 \def\Pu      {\ensuremath{u}\xspace}

 \def\thebaroffset{0.18em}
}
\newcommand{\offsetoverline}[2][\thebaroffset]{\kern #1\overline{\kern -#1 #2}}%

\makeatletter
\ifcase \@ptsize \relax
  \newcommand{\miniscule}{\@setfontsize\miniscule{4}{5}}
\or
  \newcommand{\miniscule}{\@setfontsize\miniscule{5}{6}}
\or
  \newcommand{\miniscule}{\@setfontsize\miniscule{5}{6}}
\fi
\makeatother

\DeclareRobustCommand{\optbar}[1]{\shortstack{{\miniscule (\rule[.5ex]{1.25em}{.18mm})}
  \\ [-.7ex] $#1$}}

\def\epm        {{\ensuremath{\Pe^\pm}}\xspace} 
 
\def\epem       {{\ensuremath{\Pe^+\Pe^-}}\xspace}

\def\mup        {{\ensuremath{\Pmu^+}}\xspace}
\def\mun        {{\ensuremath{\Pmu^-}}\xspace} 
 
\def\mump       {{\ensuremath{\Pmu^\mp}}\xspace} 
\def\mumu       {{\ensuremath{\Pmu^+\Pmu^-}}\xspace}

\def\taum       {{\ensuremath{\Ptau^-}}\xspace}

\def\neu        {{\ensuremath{\Pnu}}\xspace}
\def\neub       {{\ensuremath{\overline{\Pnu}}}\xspace}
\def\neum       {{\ensuremath{\neu_\mu}}\xspace}
\def\neumb      {{\ensuremath{\neub_\mu}}\xspace}

\def\neutb      {{\ensuremath{\neub_\tau}}\xspace}

\def\g      {{\ensuremath{\Pgamma}}\xspace}

\def\uquark    {{\ensuremath{\Pu}}\xspace}

\def\dquark    {{\ensuremath{\Pd}}\xspace}

\def\squark    {{\ensuremath{\Ps}}\xspace}

\def\cquark    {{\ensuremath{\Pc}}\xspace}
\def\cquarkbar {{\ensuremath{\overline \cquark}}\xspace}

\def\bquark    {{\ensuremath{\Pb}}\xspace}
\def\bquarkbar {{\ensuremath{\overline \bquark}}\xspace}

\def\tquark    {{\ensuremath{\Pt}}\xspace}
\def\tquarkbar {{\ensuremath{\overline \tquark}}\xspace}

\def\pion   {{\ensuremath{\Ppi}}\xspace}

\def\pip    {{\ensuremath{\pion^+}}\xspace}
\def\pim    {{\ensuremath{\pion^-}}\xspace}

\def\kaon    {{\ensuremath{\PK}}\xspace}
\def\Kbar    {{\ensuremath{\offsetoverline{\PK}}}\xspace}

\def\KorKbar {\kern \thebaroffset\optbar{\kern -\thebaroffset \PK}{}\xspace}

\def\Kp      {{\ensuremath{\kaon^+}}\xspace}
\def\Km      {{\ensuremath{\kaon^-}}\xspace}

\def\KS      {{\ensuremath{\kaon^0_{\mathrm{S}}}}\xspace}

\def\Kstarz  {{\ensuremath{\kaon^{*0}}}\xspace}
\def\Kstarzb {{\ensuremath{\Kbar{}^{*0}}}\xspace}
\def\Kstar   {{\ensuremath{\kaon^*}}\xspace}

\def\Dbar    {{\ensuremath{\offsetoverline{\PD}}}\xspace}
\def\D       {{\ensuremath{\PD}}\xspace}

\def\DorDbar {\kern \thebaroffset\optbar{\kern -\thebaroffset \PD}\xspace}
\def\Dz      {{\ensuremath{\D^0}}\xspace}
\def\Dzb     {{\ensuremath{\Dbar{}^0}}\xspace}
\def\Dp      {{\ensuremath{\D^+}}\xspace}
\def\Dm      {{\ensuremath{\D^-}}\xspace}

\def\DpDm    {\ensuremath{\Dp {\kern -0.16em \Dm}}\xspace}

\def\B       {{\ensuremath{\PB}}\xspace}
\def\Bbar    {{\ensuremath{\offsetoverline{\PB}}}\xspace}

\def\BorBbar {\kern \thebaroffset\optbar{\kern -\thebaroffset \PB}\xspace}
\def\Bz      {{\ensuremath{\B^0}}\xspace}
\def\Bzb     {{\ensuremath{\Bbar{}^0}}\xspace}
\def\Bd      {{\ensuremath{\B^0}}\xspace}

\def\BdorBdbar {\kern \thebaroffset\optbar{\kern -\thebaroffset \Bd}\xspace}
\def\Bu      {{\ensuremath{\B^+}}\xspace}
\def\Bub     {{\ensuremath{\B^-}}\xspace}
\def\Bp      {{\ensuremath{\Bu}}\xspace}
\def\Bm      {{\ensuremath{\Bub}}\xspace}

\def\Bs      {{\ensuremath{\B^0_\squark}}\xspace}
\def\Bsb     {{\ensuremath{\Bbar{}^0_\squark}}\xspace}
\def\BsorBsbar {\kern \thebaroffset\optbar{\kern -\thebaroffset \Bs}\xspace}
\def\Bc      {{\ensuremath{\B_\cquark^+}}\xspace}

\def\Bds     {{\ensuremath{\B_{(\squark)}^0}}\xspace}
\def\Bdsb    {{\ensuremath{\Bbar{}_{(\squark)}^0}}\xspace}
\def\BdorBs  {\Bds}

\def\jpsi     {{\ensuremath{{\PJ\mskip -3mu/\mskip -2mu\Ppsi}}}\xspace}

\def\Y#1S{\ensuremath{\PUpsilon{(#1S)}}\xspace}

\def\proton      {{\ensuremath{\Pp}}\xspace}
\def\antiproton  {{\ensuremath{\overline \proton}}\xspace}

\def\Lz          {{\ensuremath{\PLambda}}\xspace}

\def\LorLbar     {\kern \thebaroffset\optbar{\kern -\thebaroffset \PLambda}\xspace}

\def\Xires       {{\ensuremath{\PXi}}\xspace}

\def\Lc          {{\ensuremath{\Lz^+_\cquark}}\xspace}

\def\Lb           {{\ensuremath{\Lz^0_\bquark}}\xspace}

\def\BF         {{\ensuremath{\mathcal{B}}}\xspace}

\newcommand{\decay}[2]{\ensuremath{#1\!\to #2}\xspace} 

\def\to                 {\ensuremath{\rightarrow}\xspace}

\def\CP                {{\ensuremath{C\!P}}\xspace}

\def\Vub  {{\ensuremath{V_{\uquark\bquark}^{\phantom{\ast}}}}\xspace}
\def\Vcb  {{\ensuremath{V_{\cquark\bquark}^{\phantom{\ast}}}}\xspace}

\newcommand{\dms}{{\ensuremath{\Delta m_{\squark}}}\xspace}
\newcommand{\dmd}{{\ensuremath{\Delta m_{\dquark}}}\xspace}

\newcommand{\phis}{{\ensuremath{\phi_{\squark}}}\xspace}

\def\AT#1     {\ensuremath{A_{\mathrm{T}}^{#1}}\xspace}           

\def\C#1      {\ensuremath{\mathcal{C}_{#1}}\xspace}                       
\def\Cp#1     {\ensuremath{\mathcal{C}_{#1}^{'}}\xspace}                    
\def\Ceff#1   {\ensuremath{\mathcal{C}_{#1}^{\mathrm{(eff)}}}\xspace}        
\def\Cpeff#1  {\ensuremath{\mathcal{C}_{#1}^{'\mathrm{(eff)}}}\xspace}       
\def\Ope#1    {\ensuremath{\mathcal{O}_{#1}}\xspace}                       
\def\Opep#1   {\ensuremath{\mathcal{O}_{#1}^{'}}\xspace}                    

\newcommand{\aunit}[1]{\ensuremath{\text{\,#1}}}       

\newcommand{\tev}{\aunit{Te\kern -0.1em V}\xspace}
\newcommand{\gev}{\aunit{Ge\kern -0.1em V}\xspace}
\newcommand{\mev}{\aunit{Me\kern -0.1em V}\xspace}
\newcommand{\kev}{\aunit{ke\kern -0.1em V}\xspace}
\newcommand{\ev}{\aunit{e\kern -0.1em V}\xspace}
 
\newcommand{\mevc}{\ensuremath{\aunit{Me\kern -0.1em V\!/}c}\xspace}
\newcommand{\gevc}{\ensuremath{\aunit{Ge\kern -0.1em V\!/}c}\xspace}
\newcommand{\mevcc}{\ensuremath{\aunit{Me\kern -0.1em V\!/}c^2}\xspace}
\newcommand{\gevcc}{\ensuremath{\aunit{Ge\kern -0.1em V\!/}c^2}\xspace}

\def\cm   {\aunit{cm}\xspace}

\def\fb   {\ensuremath{\aunit{fb}}\xspace}
\def\invfb   {\ensuremath{\fb^{-1}}\xspace}

\def\sec  {\ensuremath{\aunit{s}}\xspace}

\def\ps   {\ensuremath{\aunit{ps}}\xspace}

\def\gsim{{~\raise.15em\hbox{$>$}\kern-.85em
          \lower.35em\hbox{$\sim$}~}\xspace}
\def\lsim{{~\raise.15em\hbox{$<$}\kern-.85em
          \lower.35em\hbox{$\sim$}~}\xspace}

\def\pt         {\ensuremath{p_{\mathrm{T}}}\xspace}

\def\degrees{\ensuremath{^{\circ}}\xspace}

\def\mrad{\aunit{mrad}\xspace}

\def\tell1  {TELL1\xspace}
\def\ukl1   {UKL1\xspace}

\newcommand{\eg}{\mbox{\itshape e.g.}\xspace}
\newcommand{\ie}{\mbox{\itshape i.e.}\xspace}

\newcommand{\etc}{\mbox{\itshape etc.}\xspace}

\newcommand{\lhcborcid}[1]{\href{https://orcid.org/#1}{\hspace*{0.1em}\raisebox{-0.45ex}{\includegraphics[width=1em]{figs/orcidIcon.pdf}}}}

\usepackage{cite} 
\usepackage{mciteplus}

\usepackage{comment}
\usepackage{tikz}
\usetikzlibrary{shapes}

\begin{document}

\renewcommand{\thefootnote}{\fnsymbol{footnote}}
\setcounter{footnote}{1}

\begin{titlepage}
\pagenumbering{roman}

\vspace*{-1.5cm}
\centerline{\large EUROPEAN ORGANIZATION FOR NUCLEAR RESEARCH (CERN)}
\vspace*{1.5cm}
\noindent
\begin{tabular*}{\linewidth}{lc@{\extracolsep{\fill}}r@{\extracolsep{0pt}}}
\ifthenelse{\boolean{pdflatex}}
{\vspace*{-1.5cm}\mbox{\!\!\!\includegraphics[width=.14\textwidth]{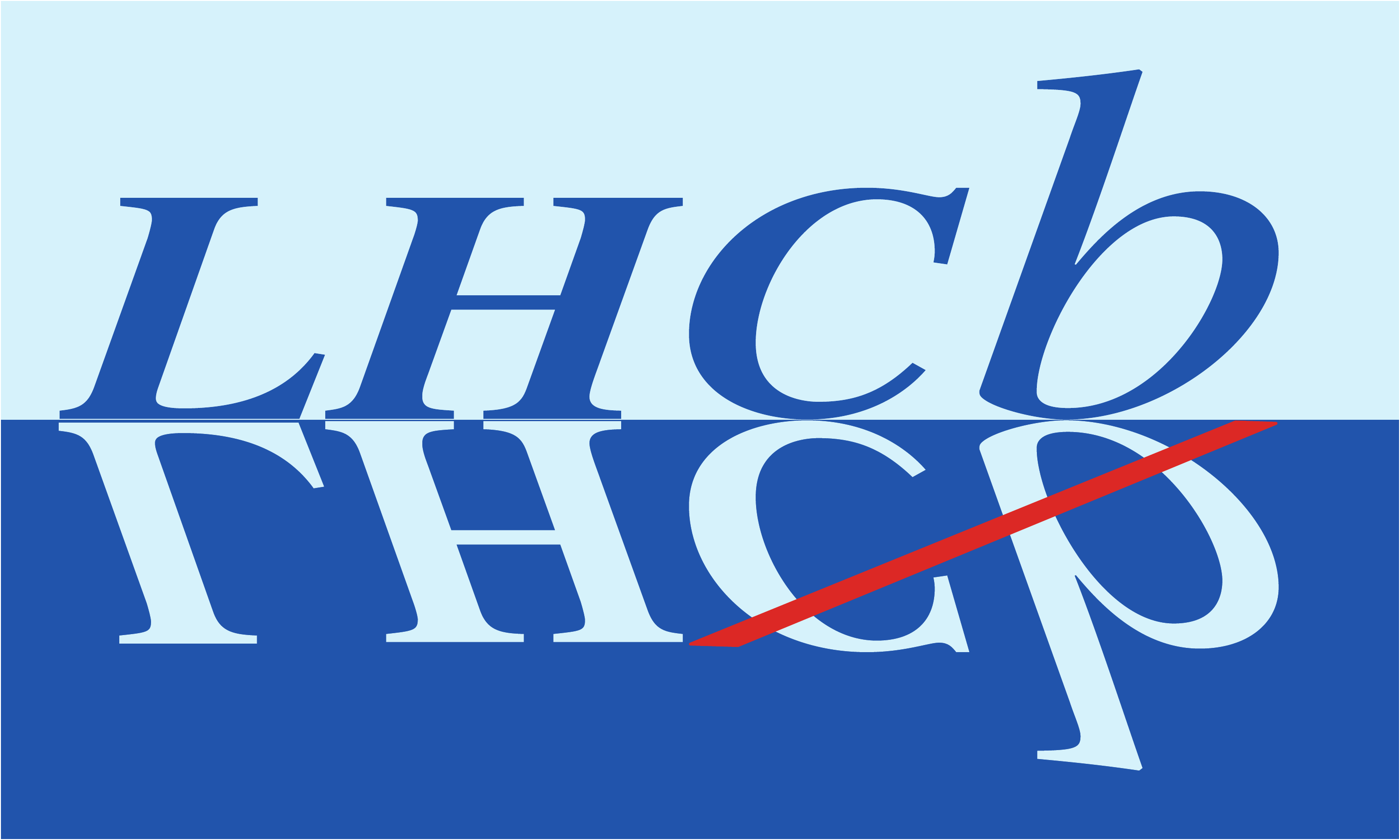}} & &}%
{\vspace*{-1.2cm}\mbox{\!\!\!\includegraphics[width=.12\textwidth]{figs/lhcb-logo.eps}} & &}%
\\
 & & LHCb-PUB-2025-001 \\  
 & & \today \\ 
 & & \\
\end{tabular*}

\vspace*{2.0cm}

{\normalfont\bfseries\boldmath\huge 
\begin{center}
    \papertitle \\
    \vspace*{0.5cm}
  {\normalsize Input to the European Particle Physics Strategy Update 2024--26}
\end{center}
}

\vspace*{1.0cm}

\begin{center}
\paperauthors\footnote{
    Contact authors: 
    Vincenzo Vagnoni (\href{mailto:vincenzo.vagnoni@cern.ch}{vincenzo.vagnoni@cern.ch}),
    Tim Gershon (\href{mailto:tim.gershon@cern.ch}{tim.gershon@cern.ch}),
    Giovanni Punzi (\href{mailto:giovanni.punzi@cern.ch}{giovanni.punzi@cern.ch})
}
\end{center}

\vspace{\fill}

\begin{abstract}
    \noindent
    A second major upgrade of the LHCb detector is necessary to allow full exploitation of the HL-LHC for flavour physics.
    The new detector will be installed during long shutdown 4 (LS4), and will operate at instantaneous luminosity up to $1.5\times 10^{34}\cm^{-2}\sec^{-1}$.
    By upgrading all subsystems and adding new detection capability it will be possible to accumulate a sample corresponding to an integrated luminosity of at least $300\invfb$ of high energy $pp$ collision data, giving unprecedented and unique scientific opportunities in flavour physics, in electroweak physics, in searches for new feebly interacting particles and in hadron spectroscopy.
    In this document, the potential of the LHCb Upgrade~II detector to enable major discoveries through increased sensitivity to a range of as-yet unknown phenomena is summarised. 
\end{abstract}

\vspace{\fill}

{\footnotesize 
\centerline{\copyright~\papercopyright. \href{\paperlicenceurl}{\paperlicence}.}}
\vspace*{2mm}

\end{titlepage}


\newpage
\setcounter{page}{2}
\mbox{~}

\renewcommand{\thefootnote}{\arabic{footnote}}
\setcounter{footnote}{0}

\cleardoublepage


\pagestyle{plain} 
\setcounter{page}{1}
\pagenumbering{arabic}


\section{Introduction}

The successful operation and rich physics harvest of LHCb during Run~1 and Run~2 of the LHC has vindicated the concept and design of a dedicated heavy-flavour physics experiment at a hadron collider. 
This has inspired efforts to upgrade the experiment's capabilities, to operate at higher instantaneous luminosities and more fully exploit the unique physics opportunities offered by the unprecedented rates of heavy quarks that LHC collisions produce.
While the initial LHCb experiment aimed at an instantaneous $pp$ collision luminosity of $2\times 10^{32}\cm^{-2}\sec^{-1}$, the Upgrade~I of the detector is now operating at its design luminosity of $2\times 10^{33}\cm^{-2}\sec^{-1}$.
The LHCb collaboration is now preparing for another order of magnitude increase with its Upgrade~II project, aiming at luminosities in excess of $10^{34}\cm^{-2}\sec^{-1}$ and an integrated sample size corresponding to at least $300 \invfb$. 
Detailed plans for LHCb Upgrade~II have been set out in a Framework Technical Design Report~\cite{LHCb-TDR-023}, supplemented by a Scoping Document~\cite{LHCb-TDR-026}.
The physics programme is described at length in a separate document~\cite{LHCb-PII-Physics}, where more details beyond those presented here can be found.

This document provides a summary of the discovery potential of LHCb Upgrade~II, with a focus on physics beyond the Standard Model (SM), referred to as ``New Physics'' (NP).
The unique capability to discover a range of new hadrons, including exotic tetraquark and pentaquark states, is also briefly described.
Separate documents submitted to the European Particle Physics Strategy Update 2024--26 describe the technology developments required for the experiment~\cite{LHCb-PUB-2025-002}, the heavy-ion physics programme~\cite{LHCb-PUB-2025-003}, and the plans for software and offline computing~\cite{LHCb-PUB-2025-004}.

\section{Flavour physics}

Flavour physics provides unique probes of physics beyond the SM, which may be able to explain some of the most fundamental mysteries in Nature: the origin of matter-antimatter asymmetry and the questions of why there are so many fundamental fermions, and what explains the patterns in their masses and mixing parameters.
In the SM, there are no flavour-changing neutral currents at tree level, and the charged-current weak interactions have a distinctive structure including a single source of \CP violation, encoded within the Cabibbo--Kobayashi--Maskawa (CKM) quark-mixing matrix.
Thus, theoretically clean predictions can be made for many rare and \CP-violating processes in the SM, and can be compared to experimental measurements to test for NP contributions.  
Since these contributions can be due to virtual (or ``off-shell'') particles, precision measurements in the flavour sector can probe far beyond the currently achievable energy frontier with sensitivity to NP scales of $10\tev$ or higher, depending on the details of the dynamics involved.
In addition to offering unique NP discovery opportunities, the flavour sector can thus provide guidance for the strategy of future energy frontier searches.

Since the flavour physics programme of LHCb is based on searches for the quantum imprints of virtual particles, it is limited only by precision.
Numerous observables in both rare decays and \CP\ violation can be predicted in the SM with theoretical uncertainty that is either negligible or controllable using data.
Moreover, the theoretical precision is constantly advancing, thanks to the efforts of phenomenologists and lattice QCD collaborations around the world.
Therefore, the focus is on improving the experimental precision and investigating unexplored regions of phase space.
The capability to pursue this programme depends on the relevant production cross-sections, the sample size (integrated luminosity), the acceptance and selection efficiency, and the ability to suppress backgrounds and resolve relevant signatures of the signal.  
The first two runs of the LHC have established LHCb as the world's leading flavour physics experiment, demonstrating the necessary experimental capability to exploit the enormous cross-sections for production of charm and beauty hadrons in LHC collisions.
However, some of the most interesting, and theoretically clean, observables remain unexplored, while many others are still statistically limited with experimental uncertainties considerably larger than those of the predictions of the models being tested.
A step-change in experimental sensitivity will be achieved at LHCb Upgrade~II by increasing the rate at which data are collected and processed, while maintaining or improving the performance of the detector in terms of signal efficiencies, background rejection and other analysis performance metrics.
In particular, the flavour physics programme requires the precision vertexing information provided by the Vertex Locator (VELO), high resolution momentum determination from the tracking system, and the capability to positively identify both electrons and muons and to distinguish between different charged hadrons. 
The higher number of $pp$ collisions per bunch-crossing in LHCb Upgrade~II makes precise timing capability necessary in order to suppress backgrounds to comparable levels to those in the current experiment.

\paragraph{Standard Model benchmarks.}
The CKM quark-mixing matrix provides the only known source of \CP violation. 
Matter-antimatter asymmetries arising from this source are observable only in a limited and clearly predicted set of decay rates of hadrons, the majority of which are in the heavy-flavour sector.
The small size of \CP violation in the SM is insufficient, by several orders of magnitude, to explain the observed matter-antimatter asymmetry of the Universe.
Consequently, there must be additional sources of \CP violation in nature beyond the SM.
Moreover, many NP models provide new \CP-violating interactions. 
This motivates searches for \CP-violating NP in as wide a range of processes as possible.

\begin{figure}[!tb]
    \centering
    \includegraphics[width=0.49\textwidth]{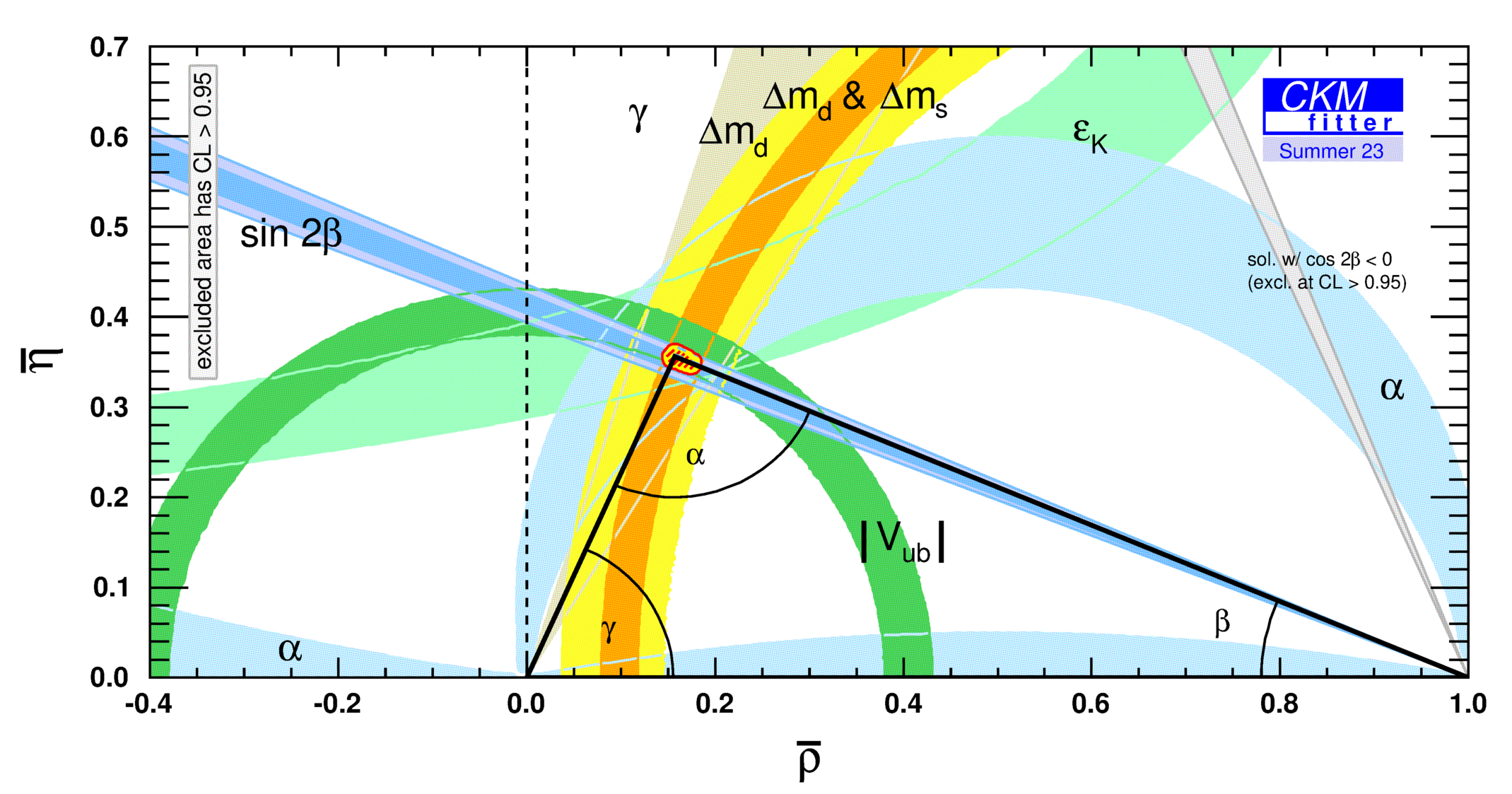}
    \includegraphics[width=0.47\textwidth]{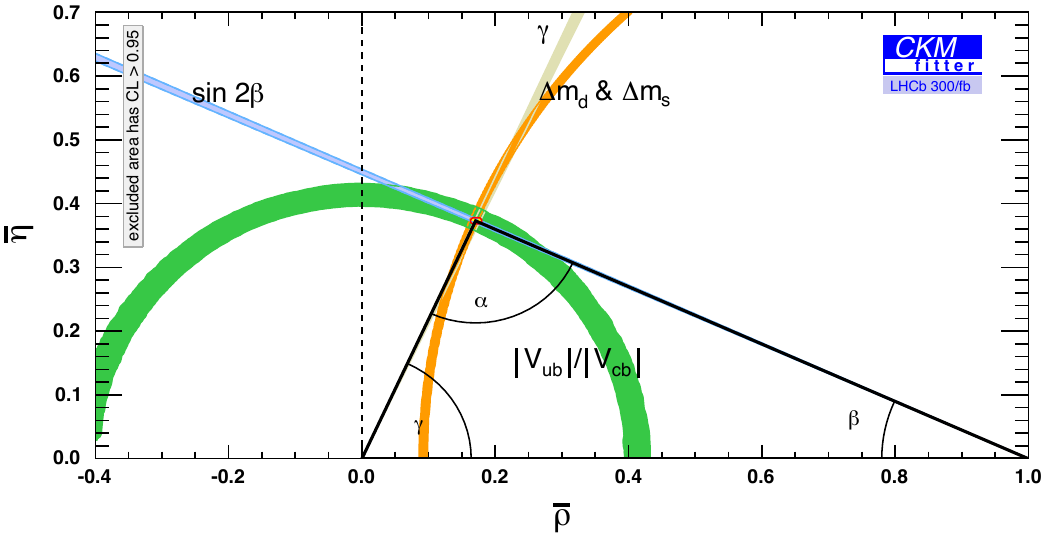}
    \caption{Constraints from the dominant CKM observables to the apex of the unitarity triangle $(\bar{\rho},\bar{\eta})$ with (left)~global inputs as of 2023~\cite{CKMfitter2005} and (right)~LHCb Upgrade~II measurements with a sample corresponding to an integrated luminosity of $300\invfb$ and anticipated improvements in lattice QCD calculations~\cite{LHCb-PII-Physics}.}
    \label{fig:ckm}
\end{figure}

Searches for \CP-violating NP require precise determination of SM benchmarks, \ie\ measurements that can be related to the fundamental SM parameters with minimal theoretical uncertainty.
The total amount of \CP violation in the quark sector of the SM is governed by the position of the apex of the CKM unitarity triangle, which can be constrained using several complementary measurements.
As shown in Fig.~\ref{fig:ckm}, the apex can be determined using only tree-level processes (CKM angle \g and magnitudes of CKM elements \Vub and \Vcb) or instead using flavour-changing neutral-current (FCNC) loop-level processes (CKM angle $\beta$ and measurements of neutral $B$-meson oscillation rates, \dmd and \dms).
By over-constraining the CKM model, \CP-violating NP can become manifest through tension between tree- and loop-level determinations.

The golden SM benchmark is the CKM angle \g, which can be determined with negligible theoretical uncertainty entirely from tree-level processes such as $\Bm\to D\Km$ decays.
Several complementary methods for determining \g involve different intermediate neutral $D$ meson decays, and their dominant systematic uncertainties arise from different sources.
This provides robustness against systematic uncertainties that may affect particular analyses; current estimates of the relevant effects indicate that systematic uncertainties will remain subdominant even with the full LHCb Upgrade~II sample size.

The latest data from LHCb give a measurement of \g with a precision better than~$3\degrees$.
This can be compared to the precision obtained by predicting the value from other measurements of CKM matrix elements; such predictions currently have precision of~$\sim 1\degrees$, which will improve further as other measurements and theoretical calculations are refined. 
With LHCb Upgrade~II the precision of the ``direct'' determination from $\Bm\to D\Km$ and similar decays will be improved down to $\sim 0.3\degrees$.
This will make \g the most precisely determined SM benchmark of the CKM paradigm against which all other CKM observables can be compared.

The value of $\sin 2\beta$ can be determined from the decay-time-dependent \CP asymmetry in $\Bz \to \jpsi \KS$ decays in what is usually considered a theoretically clean manner.
The LHCb measurement of this quantity from Run~1 and~2 data is the world's most precise and remains statistically limited, demonstrating that significant further reduction in uncertainty can be anticipated with LHCb Upgrade~II.
At that level of precision, however, interpretation of the measured \CP\ parameter in terms of $\sin 2\beta$ will be limited by theoretical uncertainties due to subleading SM amplitudes.
These can be reduced exploiting flavour symmetry relations.
For this, LHCb's capability to study decays of both $\Bz$ and $\Bs$ mesons to a wide range of final states will be crucial.
Obtaining the best constraint on the apex of the unitarity triangle also requires improvements in theory, for example the lattice QCD calculations that are needed to convert the already very precise measurements of $\Delta m_d$ and $\Delta m_s$ into bounds in the $(\bar{\rho},\bar{\eta})$ plane.

There are also excellent prospects for measurements of $\left|\Vub\right|$ and $\left|\Vcb\right|$ with the Upgrade~II detector.
Several modes that are currently inaccessible, \eg those involving decays of \Bc mesons, will become measurable with the large Upgrade~II dataset.
Furthermore, the planned detector improvements will greatly enhance opportunities for $\left|\Vub\right|$ extraction with the $\mbox{\decay{\Bsb}{\Kp\mun\neum}}$ and $\mbox{\decay{\Lb}{\proton\mun\neumb}}$ decays.
The proposed thinning of the shield that separates the VELO sensors from the LHC beams will improve significantly the resolution of the corrected mass variable that is vital to distinguish signal from background in modes with a neutrino in the final state.
The introduction of the time-of-flight TORCH detector will provide accurate identification of the low momentum particles that can arise in important phase-space regions of these decays, complementing the kinematic range of LHCb's Ring Imaging Cherenkov (RICH) detectors.

A comparison of the current LHCb CKM constraints with the predicted Upgrade~II sensitivity can be seen in Fig.~\ref{fig:ckm}, showing the unprecedented precision that can be reached.
In this plot it is assumed that only SM amplitudes contribute, so that all constraints overlap at a common point, namely the apex of the unitarity triangle.  
Once the measurements are made, whether the constraints overlap or not will allow either strong constraints on, or the discovery of, NP contributions.

\paragraph{\texorpdfstring{\boldmath New Physics in \CP violation.}{New Physics in CP violation.}}
Generic NP models often provide new sources of \CP violation which could explain the large discrepancy between the observed matter--antimatter asymmetry of the Universe and the amount that can be generated through SM processes in the early Universe.
New sources of \CP violation could affect some of the SM benchmarks discussed above, becoming visible through inconsistency of the unitarity triangle fits.  
There are also many other attractive approaches to search for new sources of \CP violation that can be pursued at LHCb Upgrade~II.

The \CP-violating weak phase associated to $\Bs$--$\Bsb$ oscillations, \phis, is a particularly sensitive probe of NP models as it is both extremely small and very precisely predicted in the SM, so that subtle NP contributions can be detected.
The SM prediction, $\phis = 37\pm 1 \mrad$, comes from the SM benchmarks mentioned in the previous section.
The latest precision measurements of $\phis$ by LHCb and CMS are approaching the sensitivity needed to observe a nonzero value, and this milestone may be achieved with data available before the end of Run~4.
The improvement in uncertainty to the ${\cal O}(1\mrad)$ level made possible by Upgrade~II will provide the ultimate test of compatibility of this phase with its SM prediction.  
Importantly, this includes studies of a number of different $\bquark\to\cquark\cquarkbar\squark$ decay modes, including polarisation-dependent measurements in $\Bs \to \jpsi \phi$ decays, as well as processes related by flavour symmetries.
This will allow the origin of a deviation from the prediction to be disentangled as being due to either NP or a subleading SM amplitude. 

The phase \phis can also be extracted from decays to final states that proceed only through loop processes such as $\Bs\to\phi\phi$ and $\Bs\to\Kstarz\Kstarzb$.
These decays are highly sensitive to NP, since virtual particles contribute to both the mixing and decay amplitudes.
Furthermore, flavour symmetry relations between $\Bs\to\Kstarz\Kstarzb$ and $\Bd\to\Kstarz\Kstarzb$ can be used to constrain precisely contributions from subleading SM amplitudes, greatly reducing the theoretical uncertainty in interpretation of the results in those modes.  

Violation of \CP symmetry in both $\Bz\text{--}\Bzb$ and $\Bs\text{--}\Bsb$ mixing is also expected to be extremely small in the SM, and therefore provides an excellent null test through which NP can be searched for.
The measurements are typically made using semileptonic decays, with observables denoted $a_{\rm sl}^{d(s)}$ for the $\Bds\text{--}\Bdsb$ systems.
The existing LHCb results are already world-leading, and a significant improvement in sensitivity can be achieved with Upgrade~II.

\paragraph{New Physics in charm.} 
Charm hadrons provide a unique opportunity to study \CP violation in FCNC transitions involving up-type quarks.
These can be affected by NP contributions in fundamentally different ways to the down-type quarks in the kaon and beauty systems.
Since the level of \CP violation expected in the charm system is extremely small, $\mathcal{O}(10^{-4})$, it provides a potentially very sensitive NP probe, but uncertainties related to long-distance QCD interactions limit the precision of the theoretical predictions.
LHCb has made the first observation of \CP violation in charm decays with a measurement of the asymmetry $\Delta A_{\CP} = A_{\CP}(\Dz\to\Kp\Km) - A_{\CP}(\Dz\to\pip\pim)$ consistent with, but at the top end of, the range of SM predictions.
Further measurements with other processes are necessary to understand whether the observed \CP\ violation can be explained within the SM.

In particular, measurements of \CP\ violation associated to $\Dz\text{--}\Dzb$ mixing, which are expected to be at the $\mathcal{O}(10^{-5})$ level, are crucial. 
The nonzero value of $\Delta A_{\CP}$ shows clearly that there are two amplitudes with different weak and strong phases contributing to at least one of the $\Dz\to\Kp\Km$ and $\Dz\to\pip\pim$ decays. 
Determination of individual \CP\ asymmetries helps to pin down where the effect comes from.
Still, further input is essential to understand to what extent each \CP\ asymmetry is driven by a larger-than-expected ratio of the magnitudes of the two amplitudes or by the strong phase difference, and thus to establish whether the observed value of $\Delta A_{\CP}$ can be explained by the SM or not. 
Determination of decay-time-dependent asymmetries in $\Dz\to\Kp\Km$ and $\Dz\to\pip\pim$ decays will provide this input, but only if sensitivity at the level of $10^{-5}$ or better can be achieved.
This extreme level of precision is necessary since the interesting effects are suppressed by the small values of the mixing parameters in the $\Dz$--$\Dzb$ system.
The current measurements have precision at the $10^{-4}$ level and are statistically limited, showing that this target is within reach as long as LHCb Upgrade~II collects $300 \invfb$ while maintaining detector performance at the current levels.
Improvements in the acceptance for low momentum charged particles, through the introduction of Magnet Stations instrumenting the sides of LHCb's dipole magnet, will allow reduction of the uncertainty beyond that from luminosity scaling alone.
LHCb Upgrade~II is the only experiment, existing or planned, that can achieve the necessary precision to understand whether charm \CP\ violation is caused by NP.

\paragraph{New Physics in rare decays.}
The nonexistence of tree-level FCNC transitions is a feature that is highly specific to the SM.
There is no fundamental necessity for the cancellation of these processes, and consequently generic NP models often provide sources of FCNCs.
Decays which can proceed only through FCNCs are therefore highly sensitive probes of NP, as the NP amplitudes are potentially large compared to the small SM contribution.

Historically, the golden channel for NP searches in FCNC $b$-hadron decays has been $\Bs\to\mumu$.
LHCb was the first experiment to observe independently this extremely rare ($\sim 10^{-9}$) decay and in the latest analysis a hint of the even-further suppressed ($30$ times rarer) mode $\Bd\to\mumu$ appears to be emerging.
Together, measurements of these two branching fractions provide extremely powerful tests of the SM with ability to discriminate between different NP models, in particular testing the minimal flavour violation NP scenario.
With the Upgrade~II dataset, LHCb will additionally have capability to measure the parameters $A_{\Delta\Gamma}^{\mu\mu}$ and $S_{\mu\mu}$ of the $\Bs\to\mumu$ decay-time distribution.
These are considered smoking gun observables that, if different from their SM expectations of unity and zero respectively, would provide unambiguous evidence for~NP.

LHCb has made many world-first and world-best measurements in flavour-changing \mbox{\decay{\bquark}{\squark\ell^+\ell^-}} and \mbox{\decay{\bquark}{\dquark\ell^+\ell^-}} transitions.
These have revealed a consistent pattern of deviations from SM predictions in branching fractions (\eg~${\cal B}\left(\Bp\to\Kp\mumu\right)$, ${\cal B}\left(\Bz\to\Kstarz\mumu\right)$, ${\cal B}\left(\Bs\to\phi\mumu\right)$, ${\cal B}\left(\Lb\to\Lz\mumu\right)$) and angular observables (\eg~the $P_5^\prime$ observable in $\Bz\to\Kstarz\mumu$ decays), often collectively referred to as the $B$ anomalies.
With the current data it is not possible to determine conclusively whether these arise from larger than expected QCD effects or from NP, but several data-based methods have been proposed that can help to shed light on this question.  
First implementations of these approaches with existing data are statistically limited; more definitive answers will be obtained with the LHCb Upgrade~II data sample.
Additionally, the size and quality of the sample will allow the precision of \mbox{\decay{\bquark}{\dquark\ell^+\ell^-}} studies to reach a similar level to that currently achieved for \mbox{\decay{\bquark}{\squark\ell^+\ell^-}} transitions.

\begin{figure}[!tb]
    \centering
    \includegraphics[width=0.75\textwidth]{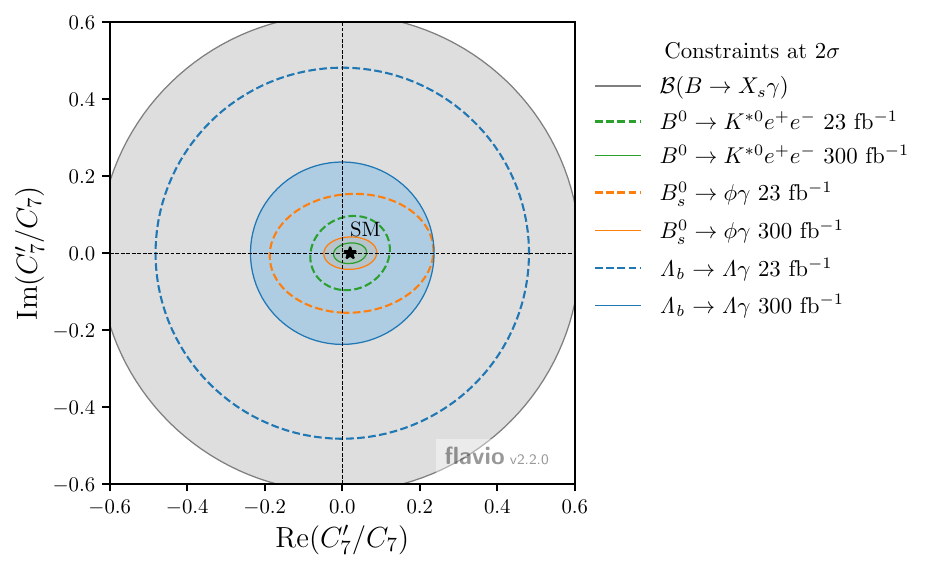}
    \caption{Constraints that can be obtained by the end of Run~3 and at the completion of LHCb Upgrade~II on the real and imaginary parts of the Wilson coefficient $C_7^\prime$ that quantifies the strength of right-handed $b \to s \gamma$ interactions from each of $\Bz\to\Kstarz\epem$, $\Bs \to \phi \gamma$ and $\Lb \to \Lz\gamma$ decays~\cite{LHCb-PII-Physics}.
    The SM expectation of negligible right-handed currents is also shown.}
    \label{fig:C7prime}
\end{figure}

The chiral structure of the weak interaction is a further distinctive feature of the SM that is not necessarily replicated in NP scenarios.  
New right-handed currents that appear at high energy scales could, through quantum-loop effects, leave an imprint on the polarisation of the photon emitted in $b \to s \gamma$ processes.  
This can be probed through several different methods, including studies of the angular distribution in $\Bz\to\Kstarz\epem$ decays at low $\epem$ invariant mass, studies of the decay-time distribution in $\Bs \to \phi \gamma$ decays, and studies of the angular distributions in $\Lb \to \Lz\gamma$ decays.
The last two of these methods are unique to LHCb, and with the Upgrade~II data sample and PicoCal electromagnetic calorimeter the former will also be measured more precisely than at any other experiment.  
The precision that can be achieved, in terms of the Wilson coefficient $C_7^\prime$ that quantifies the strength of right-handed $b \to s \gamma$ interactions, is illustrated in Fig.~\ref{fig:C7prime}.

In addition to rare beauty decays, LHCb has unique capability to probe rare decays of charm and strange hadrons. 
For example, LHCb already has the world's best limit on the branching fraction of $\mbox{\decay{\KS}{\mumu}}$ decays, and with the Upgrade~II data sample will push the sensitivity to the $10^{-11}$ level, approaching the SM prediction.

\paragraph{New Physics in lepton flavour.}
An enticing possibility is that NP may be causing the observed \decay{\bquark}{\squark\ell^+\ell^-} deviations exclusively in \decay{\bquark}{\squark\mumu} and, perhaps, \decay{\bquark}{\dquark\mumu} processes while leaving the corresponding \decay{\bquark}{\squark\epem} and \decay{\bquark}{\dquark\epem} transitions at their SM rates.
This would provide striking violations of lepton flavour universality (LFU), which could not be caused by any SM process. 
Although the hints seen in early LHCb measurements of the branching fraction ratios $R_K$ and $R_{\Kstar}$ were not confirmed in improved analyses with larger data samples, this remains a powerful way to probe for NP in which LHCb Upgrade~II has unparalleled precision.

In addition, LFU in the charged-current \decay{\bquark}{\cquark\taum\neutb} decays will be comprehensively explored across all \bquark-hadron species at LHCb Upgrade~II. 
Current measurements of $R(D)$ and $R(D^*)$, where  \mbox{$R(H_c) = {\cal B}\left(H_b \to H_c \taum\neutb\right)/{\cal B}\left(H_b \to H_c \mun\neumb\right)$} and $H_b$ and $H_c$ represent beauty and charm hadrons respectively, are in tension with SM predictions.
Significant improvements in precision of these LFU ratios across all \bquark-hadron species will allow potential observation of NP contributions and, potentially, provide early hints as to the coupling structure.
To fully exploit these decays’ discovery potential, it is crucial to go beyond branching ratios and explore differential properties like angular distributions. 
Even if LFU ratios match the SM, NP could be observed through subtle deviations in differential rates. 
The high dimensionality of such analyses demands LHCb Upgrade~II’s vast statistics to control backgrounds with data-based methods. 
Additionally, the dataset enables analogous studies in suppressed \mbox{$\decay{\bquark}{\uquark\taum\neutb}$} transitions, testing whether potential NP contributions align with the SM’s flavour structure.

LHCb Upgrade~II will also allow precise searches for NP processes that violate lepton number conservation in processes involving different lepton families, either in leptonic decays such as \decay{\BdorBs}{\epm\mump} or semileptonic \decay{\bquark}{\squark\epm\mump} processes. 
These can arise in connection with NP that violates lepton flavour universality and can also appear in other interesting NP scenarios which do not respect the accidental charged lepton flavour symmetries of the SM. 
Limits on such processes can be improved over those obtained in the existing dataset by typically an order of magnitude. 
Similarly, the sensitivity to lepton number violating modes like \decay{\Bp}{\pim\mup\mup} and \decay{\Lc}{\antiproton\mup\mup} (without or with associated baryon number violation, respectively) can be improved upon by around an order of magnitude.
Such processes can occur in NP models with either off-shell mediators, in which case all final state particles originate from the same decay vertex, or with new light degrees of freedom.
In the latter case the new particles can be long-lived resulting in a distinctive topology for the decay, which LHCb has the ideal geometry to observe, as discussed in Sec.~\ref{sec:fips}.

\section{\texorpdfstring{\boldmath Electroweak and high-\pt physics}{Electroweak and high pT physics}}
LHCb's geometry and momentum coverage provide access to a kinematic region that is inaccessible at other LHC experiments.
This has enabled a unique programme of electroweak measurements that were not foreseen in the original design of the experiment.
In particular, muonic $W$ and $Z$ decays have been used to determine the $W$ mass and the effective weak mixing angle.
The complementary acceptance of LHCb to the ATLAS and CMS experiments is particularly important here since measurements in different kinematic regions are affected differently by uncertainties in the parton distribution functions (PDFs) of the colliding protons.  
As a consequence, determination of the $W$ mass in the forward acceptance strongly suppresses the uncertainty associated to knowledge of the PDFs in an LHC $W$ mass average. 
A $W$ mass measurement with a precision of a few~\mev will be possible with LHCb Upgrade~II alone, and will be key to obtaining the best sensitivity to this crucial SM parameter from the LHC. 

Studies of $W^+$, $W^-$ and $Z$ boson production in the LHCb acceptance, either inclusively or in association with jets, are also extremely important to constrain PDFs.  
Measurements of top quark production provide additional constraints (in addition to being of interest in their own right to test the SM and search for NP).
With the Upgrade~II dataset LHCb will be able to determine the inclusive $\tquark\tquarkbar$ production cross-section to $\sim 4\%$ precision.
Improved knowledge of PDFs will have significant impact across the LHC physics programme, for example in studies of the Higgs boson and NP searches at ATLAS and CMS.

Studies of the SM Higgs boson at LHCb have, to date, been limited by both reduced luminosity and reduced production rate within the detector acceptance relative to ATLAS and CMS.
The large data sample of LHCb Upgrade~II will, however, allow interesting sensitivity in several channels.
The $HV$ production channel ($V = W$ or $Z$) with $\decay{H}{\bquark\bquarkbar}$ decay should be observable, and will allow a test of the SM prediction for the production rate in the forward region.
In addition, the same production process with $\decay{H}{\cquark\cquarkbar}$ decay can be searched for, benefitting from excellent charm jet tagging, to achieve a highly stringent constraint corresponding to approximately twice the expected value with the SM charm Yukawa coupling.

\section{Searches for new feebly interacting particles}
\label{sec:fips}
The excellent LHCb tracking system together with the development of a fully software trigger scheme, allowing reconstruction and selection algorithms to be executed in real time, make LHCb a unique and powerful lifetime frontier experiment.
As such, LHCb is able to perform highly sensitive searches for new feebly interacting particles (FIPs), which have long lifetimes due their tiny couplings to the SM. 

\begin{figure*}[b!]
    \centering
    \includegraphics[width=0.47\textwidth]{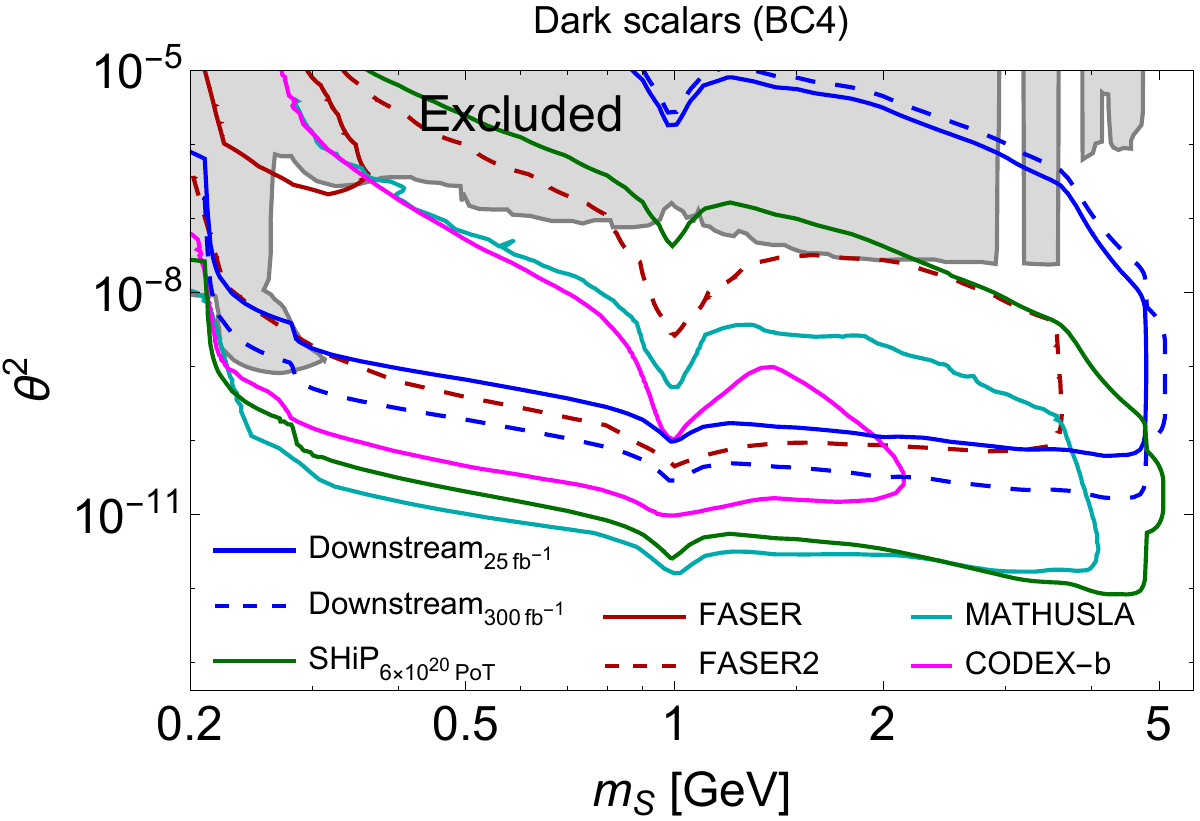}
    \includegraphics[width=0.47\textwidth]{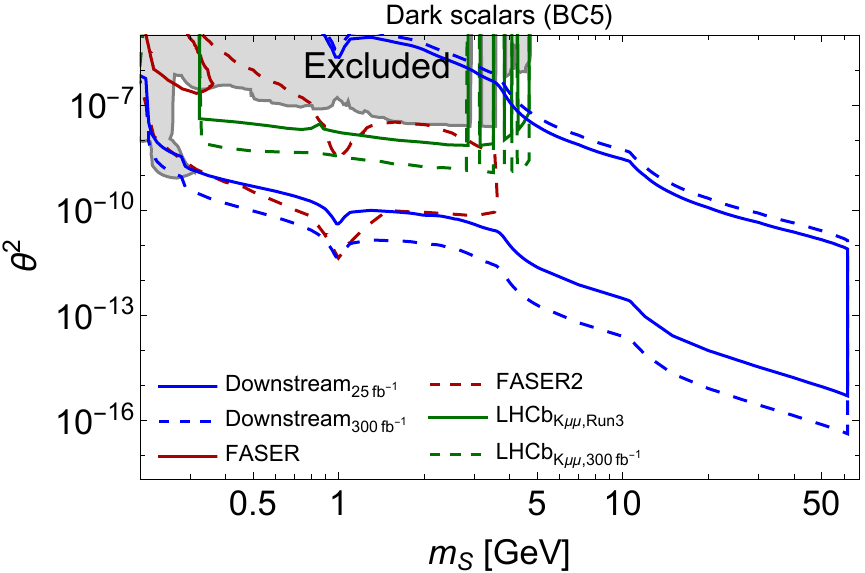}
    \caption{Expected LHCb (``Downstream'') sensitivity to the Higgs-like dark scalar model after Run~3 ($25\invfb$) and LHCb Upgrade~II ($300\invfb$), compared to other proposed lifetime-frontier experiments~\cite{Gorkavenko:2023nbk}. 
    The expected exclusion regions on the squared mixing-coupling $\theta^2$ are shown as functions of the mass of the new scalar boson $m_S$, assuming negligible background.
    Scenarios where the dark scalar is produced in (left)~$b$-hadron decays and (right) Higgs boson decays are shown.}
    \label{fig:landscape}
\end{figure*}

A wide range of models involving FIPs such as dark photons, Higgs-like dark scalars, heavy neutral leptons, and axion-like particles have been proposed as possible NP scenarios. 
Many of these models can explain the nature of dark matter, which is one of the most important unanswered questions in contemporary physics.  
These models propose new force mediators which both couple to dark matter and mix with SM particles. 
The main production modes for many \gev-scale FIPs involve $b$-hadron decays, and hence LHCb is ideally suited to search for them. 
LHCb also has excellent sensitivities to other production modes, including the case where new mediators are produced in Higgs boson decays.
Figure~\ref{fig:landscape} shows the expected sensitivity of LHCb Upgrade~II to two such models compared to other proposed lifetime-frontier experiments.  

The performance of the LHCb Upgrade~II tracking and particle identification systems, together with the new trigger and reconstruction software capabilities and the large dataset, will allow corners in phase space to be explored that have not previously been accessible.
FIPs with lifetimes above 100\ps are particularly interesting since this region of phase space is not yet excluded, and the only SM particles with similar lifetimes that can cause backgrounds to two-track signatures of FIP decays are \KS mesons and \Lz baryons.
New reconstruction and selection techniques using only trackers downstream of the LHCb dipole magnet, and including multi-track vertices, will further increase the sensitivity to a range of NP models.
For example, in the Higgs-like dark scalars model these techniques will improve the sensitivity beyond that shown in Fig.~\ref{fig:landscape}.

\section{Hadron spectroscopy}
In addition to its discovery potential for physics beyond the SM, LHCb Upgrade~II provides unique opportunities to uncover new hadronic states.
The enormous production rates for heavy quarks in LHC collisions mean that an unprecedented range of hadrons are formed, making it the ideal laboratory for spectroscopy studies.
LHCb's unique detection capability has led to a wealth of remarkable discoveries in this field, including the pentaquark ($P_{c\bar{c}}$) states and the first observations of doubly charmed hadrons, both conventional ($\Xires_{cc}$) baryons and exotic ($T_{cc}$) tetraquarks.
These discoveries have reinvigorated the field of heavy-flavour spectroscopy, prompting a new PDG naming scheme, and opened the door to a fundamental understanding of the binding mechanisms of multi-quark hadrons.

Whilst the existence of hadrons with exotic quark content was suggested in the original formulations of the quark model, there is not yet a comprehensive understanding of how these states interact and decay.
Many states appear near meson-antimeson or meson-baryon mass thresholds, suggesting a possible molecular or hadron rescattering interpretation.
However some exotic states cannot be explained in this way, indicating that models involving tight binding may play a role.
Experimental measurements of the properties of these new states are of the utmost importance because it is extremely difficult to make first-principle predictions of the excitation spectrum and widths of such multi-quark states due to the nonperturbative nature of QCD at low energies. 

With the significantly increased LHCb Upgrade~II data sample, the main limitation on which states can be observed will be due to detection capability, and therefore improved performance of the detector will have a direct impact on physics output.
Among a huge number of possible discoveries, it is anticipated that observations of many doubly heavy hadrons, both conventional and exotic, will become within reach.
This includes the doubly heavy tetraquark states ($T_{bc}$ and $T_{bb}$), at least some of which are expected to have large enough binding energies such that only weak decays are possible. 
The results of these investigations will shed light on the nature of hadronic binding mechanisms, and more generally about QCD. 
In addition to studies of prompt production, it will become possible for the first time to study production of exotic hadrons with open or hidden charm in \Bc\ meson decays, extending the mass range that is currently accessible in \Bp\ decays.  
Many other approaches that are currently inaccessible will also become within reach with the Upgrade~II dataset.

\section{Summary}
The current LHCb experiment is established as the world's leading flavour physics facility.
Results from the LHC to date continue to demonstrate that SM effectively describes phenomena accurately up to an energy scale of ${\cal O}(1\tev)$.
There must be physics beyond the SM, however, and there are strong reasons to believe that it will be accessible with the improved sensitivity that will be made possible with LHCb Upgrade~II.
The HL-LHC era offers an opportunity to collect an unprecedented data sample for dedicated heavy-flavour measurements.
In addition to the vastly increased data sample, improvements in the LHCb Upgrade~II detector will enable access to several new observables and reduce the uncertainties of other key measurements to levels comparable to their theory predictions.
The sensitivity to quantum imprints of new particles will push the NP discovery potential to energy scales of $10\tev$ and higher, far beyond what is currently achievable at the energy frontier.  
In case the NP scale is beyond the reach of ATLAS and CMS, this may be the only way that physics beyond the SM can be discovered at the HL-LHC.

The current and anticipated future uncertainties for some key flavour observables are summarised in Table~\ref{tab:uncerts}.
These are based on extrapolations from existing results with the Run~1 and~2 data, accounting for improved efficiency due to the removal of the hardware trigger since the start of Run~3. 

\begin{table}[!tb]
  \centering
  \caption{
    Uncertainties from latest LHCb measurements for some key flavour observables and projections for future upgrades~\cite{LHCb-TDR-026}.
    Upgrade~I projections are given both with the data sample available after Run~3 ($23\invfb$) and with that after Run~4 ($50\invfb$).
    Uncertainties are extrapolated assuming that systematic uncertainties will not becoming limiting (see Ref.~\cite{LHCb-PII-Physics} for further discussion).
  }
  \label{tab:uncerts}
  \resizebox{\textwidth}{!}{
  \begin{tabular}{l c c c c }
    \hline
    \multirow{1}{*}{Observable} & Current LHCb & \multicolumn{2}{c}{Upgrade~I} & \upgradetwo\ \\
    &  (up to $9\invfb$) & ($23\invfb$) & ($50\invfb$) & ($300\invfb$) \\
    \hline
    \multicolumn{5}{l}{\textbf{\underline{CKM tests}}} \\
    $\;\;\g$ ({\small $B \to DK$, \etc}) & $2.8\degrees$ & 1.3\degrees & 0.8\degrees  & 0.3\degrees \\
    $\;\;\phis$ ({\small $\Bs \to \jpsi \phi$}) & $20\mrad$ & $12\mrad$     & $8\mrad$    & $3\mrad$       \\
    $\;\;|\Vub|/|\Vcb|$ ({\small $\Lb \to p\mun\neumb$, \etc}) & $6\%$ & 3\%         & 2\%        & 1\%          \\
    \multicolumn{5}{l}{\textbf{\underline{Charm}}} \\
    $\;\;\Delta A_{\CP}$  ({\small $\Dz \to \Kp\Km, \pip\pim$})  & $29 \times 10^{-5}$ & $13\times 10^{-5}$ & $8\times 10^{-5}$ & $3.3\times 10^{-5}$ \\
    $\;\;A_\Gamma$  ({\small $\Dz \to \Kp\Km, \pip\pim$})  & $11 \times 10^{-5}$ & $5\times 10^{-5}$ & $3.2\times 10^{-5}$ & $1.2\times 10^{-5}$ \\
    $\;\;\Delta x$  ({\small $\Dz \to \KS\pip\pim$})  & $18 \times 10^{-5}$ & $6.3 \times 10^{-5}$ & $4.1 \times 10^{-5}$ & $1.6 \times 10^{-5}$ \\
    \multicolumn{5}{l}{\textbf{\underline{Rare decays}}} \\
    $\;\;\BF{(\Bd\to\mumu)}/\BF{(\Bs\to\mumu)}$\hspace{-6mm} & 69\% & 41\% & 27\% & 11\% \\
    $\;\;S_{\mu\mu}$  ({\small $\Bs\to\mumu$})     & ---    & ---         & ---        & 0.2 \\
    $\;\;A_{\rm T}^{(2)}$  ({\small $\Bz \to \Kstarz\epem$})  & 0.10 & 0.060 & 0.043 & 0.016 \\
    $\;\;S_{\phi \gamma}(\Bs\to \phi \gamma)$  & 0.32 & 0.093 & 0.062 & 0.025 \\
     $\;\;\alpha_{\gamma}(\Lb\to \Lz \gamma)$  & $^{+0.17}_{-0.29}$ & 0.148 & 0.097 & 0.038 \\
    \hline
  \end{tabular}
  }
\end{table}

\clearpage
\addcontentsline{toc}{section}{References}
\bibliographystyle{LHCb}
\bibliography{main,standard,LHCb-PAPER,LHCb-CONF,LHCb-DP,LHCb-TDR}

\ifx\mcitethebibliography\mciteundefinedmacro
\PackageError{LHCb.bst}{mciteplus.sty has not been loaded}
{This bibstyle requires the use of the mciteplus package.}\fi
\providecommand{\href}[2]{#2}
\begin{mcitethebibliography}{1}
\mciteSetBstSublistMode{n}
\mciteSetBstMaxWidthForm{subitem}{\alph{mcitesubitemcount})}
\mciteSetBstSublistLabelBeginEnd{\mcitemaxwidthsubitemform\space}
{\relax}{\relax}

\bibitem{LHCb-TDR-023}
LHCb collaboration, \ifthenelse{\boolean{articletitles}}{\emph{{LHCb Framework
  TDR for the LHCb Upgrade II Opportunities in flavour physics, and beyond, in
  the HL-LHC era}}, }{}
  \href{http://cdsweb.cern.ch/search?p=CERN-LHCC-2021-012&f=reportnumber&action_search=Search&c=LHCb}
  {CERN-LHCC-2021-012}, 2022\relax
\mciteBstWouldAddEndPuncttrue
\mciteSetBstMidEndSepPunct{\mcitedefaultmidpunct}
{\mcitedefaultendpunct}{\mcitedefaultseppunct}\relax
\EndOfBibitem
\bibitem{LHCb-TDR-026}
LHCb collaboration, \ifthenelse{\boolean{articletitles}}{\emph{{LHCb Upgrade II
  Scoping Document}}, }{}
  \href{http://cdsweb.cern.ch/search?p=CERN-LHCC-2024-010&f=reportnumber&action_search=Search&c=LHCb}
  {CERN-LHCC-2024-010}, 2025\relax
\mciteBstWouldAddEndPuncttrue
\mciteSetBstMidEndSepPunct{\mcitedefaultmidpunct}
{\mcitedefaultendpunct}{\mcitedefaultseppunct}\relax
\EndOfBibitem
\bibitem{LHCb-PII-Physics}
LHCb collaboration, \ifthenelse{\boolean{articletitles}}{\emph{{Physics case
  for an LHCb Upgrade II --- Opportunities in flavour physics, and beyond, in
  the HL-LHC era}},
  }{}\href{http://arxiv.org/abs/1808.08865}{{\normalfont\ttfamily
  arXiv:1808.08865}}\relax
\mciteBstWouldAddEndPuncttrue
\mciteSetBstMidEndSepPunct{\mcitedefaultmidpunct}
{\mcitedefaultendpunct}{\mcitedefaultseppunct}\relax
\EndOfBibitem
\bibitem{LHCb-PUB-2025-002}
LHCb collaboration, \ifthenelse{\boolean{articletitles}}{\emph{{Technology
  developments for LHCb Upgrade II}}, }{}
  \href{http://cdsweb.cern.ch/search?p=LHCb-PUB-2025-002&f=reportnumber&action_search=Search&c=LHCb+Notes}
  {LHCb-PUB-2025-002}, 2025\relax
\mciteBstWouldAddEndPuncttrue
\mciteSetBstMidEndSepPunct{\mcitedefaultmidpunct}
{\mcitedefaultendpunct}{\mcitedefaultseppunct}\relax
\EndOfBibitem
\bibitem{LHCb-PUB-2025-003}
LHCb collaboration, \ifthenelse{\boolean{articletitles}}{\emph{{Heavy ion
  physics with LHCb Upgrade II}}, }{}
  \href{http://cdsweb.cern.ch/search?p=LHCb-PUB-2025-003&f=reportnumber&action_search=Search&c=LHCb+Notes}
  {LHCb-PUB-2025-003}, 2025\relax
\mciteBstWouldAddEndPuncttrue
\mciteSetBstMidEndSepPunct{\mcitedefaultmidpunct}
{\mcitedefaultendpunct}{\mcitedefaultseppunct}\relax
\EndOfBibitem
\bibitem{LHCb-PUB-2025-004}
LHCb collaboration, \ifthenelse{\boolean{articletitles}}{\emph{{Computing and
  software for LHCb Upgrade II}}, }{}
  \href{http://cdsweb.cern.ch/search?p=LHCb-PUB-2025-004&f=reportnumber&action_search=Search&c=LHCb+Notes}
  {LHCb-PUB-2025-004}, 2025\relax
\mciteBstWouldAddEndPuncttrue
\mciteSetBstMidEndSepPunct{\mcitedefaultmidpunct}
{\mcitedefaultendpunct}{\mcitedefaultseppunct}\relax
\EndOfBibitem
\bibitem{CKMfitter2005}
CKMfitter group, J.~Charles {\em et~al.},
  \ifthenelse{\boolean{articletitles}}{\emph{{\CP violation and the CKM matrix:
  Assessing the impact of the asymmetric $B$ factories}},
  }{}\href{https://doi.org/10.1140/epjc/s2005-02169-1}{Eur.\ Phys.\ J.\
  \textbf{C41} (2005) 1},
  \href{http://arxiv.org/abs/hep-ph/0406184}{{\normalfont\ttfamily
  arXiv:hep-ph/0406184}}, {updated results and plots available at
  \href{http://ckmfitter.in2p3.fr/}{{\texttt{http://ckmfitter.in2p3.fr/}}}}\relax
\mciteBstWouldAddEndPuncttrue
\mciteSetBstMidEndSepPunct{\mcitedefaultmidpunct}
{\mcitedefaultendpunct}{\mcitedefaultseppunct}\relax
\EndOfBibitem
\bibitem{Gorkavenko:2023nbk}
V.~Gorkavenko {\em et~al.}, \ifthenelse{\boolean{articletitles}}{\emph{{LHCb
  potential to discover long-lived new physics particles with lifetimes above
  100 ps}}, }{}\href{https://doi.org/10.1140/epjc/s10052-024-12906-3}{Eur.\
  Phys.\ J.\  \textbf{C84} (2024) 608},
  \href{http://arxiv.org/abs/2312.14016}{{\normalfont\ttfamily
  arXiv:2312.14016}}\relax
\mciteBstWouldAddEndPuncttrue
\mciteSetBstMidEndSepPunct{\mcitedefaultmidpunct}
{\mcitedefaultendpunct}{\mcitedefaultseppunct}\relax
\EndOfBibitem
\end{mcitethebibliography}
 
\end{document}